\documentclass[9pt,twocolumn,twoside]{optica}
\setboolean{shortarticle}{true}
\setboolean{minireview}{false}
\usepackage{graphicx,amssymb,mathtools,bm,nicefrac,blindtext,mathptmx} 

\newcommand{\openone}{\leavevmode\hbox{\small1\normalsize\kern-.33em1}}

\newcommand{\sgn}{\operatorname{sgn}}

\title{Tempering Rayleigh's curse with PSF shaping}

\author[1]{Martin~Pa\'ur} 
\author[1]{Bohumil Stoklasa}
\author[2]{Jai~Grover} 
\author[2]{Andrej~Krzic} 
\author[3,4]{Luis~L.~S\'{a}nchez-Soto} 
\author[1]{Zden\v{e}k~Hradil}
\author[1]{Jaroslav~\v{R}eh\'a\v{c}ek} 

\affil[1]{Department of Optics,
 Palack\'y University, 17. listopadu 12, 771 46 Olomouc, 
Czech Republic}

\affil[2]{ESA---Advanced
  Concepts and Studies Office, European Space Research Technology
  Centre (ESTEC), Keplerlaan 1, Postbus 299, NL-2200AG Noordwijk,
  Netherlands}

\affil[3]{Departamento de \'Optica,
  Facultad de F\'{\i}sica, Universidad Complutense, 
28040~Madrid,  Spain} 

\affil[4]{Max-Planck-Institut f\"ur die Physik des Lichts,
  Staudtstra\ss e 2, 91058 Erlangen,
  Germany}
\affil[*]{Corresponding author: lsanchez@fis.ucm.es}

\dates{Compiled \today}

\ociscodes{(100.6640) Superresolution; (110.3055) Information
  theoretical analysis; (120.3940)   Metrology.}

\doi{\url{http://dx.doi.org/10.1364/aop.XX.XXXXXX}}

\begin{abstract}
  It has been argued that, for a spatially invariant imaging system,
  the information one can gain about the separation of two incoherent
  point sources decays quadratically to zero with decreasing
  separation, an effect termed Rayleigh's curse.  Contrary to this
  belief, we identify a class of point spread functions with a linear
  information decrease.  Moreover, we show that any well-behaved
  symmetric point spread function can be converted into such a form
  with a simple nonabsorbing signum filter. We experimentally
  demonstrate significant superresolution capabilities based on this
  idea.
\end{abstract}

\setboolean{displaycopyright}{true}

\begin{document}

\maketitle
\thispagestyle{fancy}

\ifthenelse{\boolean{shortarticle}}{\abscontent}{}

Rayleigh's criterion, which stipulates a minimum separation for
two equally bright incoherent point light sources to be
distinguishable by an imaging system, is based on heuristic notions. A
more accurate approach to optical resolution can be formulated in
terms of the Fisher information and the associated Cram\'er-Rao lower
bound (CRLB), which sets a limit on the precision with which the
separation can be estimated~\cite{Kay:1993aa}.  Using statistical
methods, it has been found that the resolution predicted by 
Rayleigh's criterion has been routinely 
improved~\cite{Dekker:1997aa,Hemmer:2012aa}.

A recent reexamination of the problem by Tsang and
coworkers~\cite{Tsang:2016aa,Nair:2016aa,Nair:2016ab,Tsang:2017aa}
showed that, for direct imaging, the Fisher information drops
quadratically with the object separation.  In consequence, the
variance of any estimator based in these intensity-based measurements
diverges, giving rise to the so-called Rayleigh's curse~\cite{Tsang:2016aa}.
Surprisingly, when one calculates the quantum Fisher
information~\cite{Petz:2011aa} (i.e., optimized over all the possible
measurements), the associated quantum CRLB maintains a constant
value. This shows that, in principle, the separation can be estimated
with precision unaffected by Rayleigh's curse. The key ingredient for
this purpose is to utilize phase-sensitive
measurements~\cite{Lupo:2016aa,Rehacek:2017aa}. This has been 
demonstrated by holographic mode projection~\cite{Paur:2016aa},
heterodyne detection~\cite{Yang:2016aa,Yang:2017aa}, and
parity-sensitive interferometers~\cite{Tham:2016aa}.

These experiments dispel Rayleigh's curse, but require sophisticated
equipment. In this Letter, we revisit the scenario of direct
detection, for it is the cut-and-dried method used in the
laboratory. We show that by using a simple phase mask---a signum
filter---direct detection makes the Fisher information drop linearly.
This scaling law opens new avenues for boosting resolution, as we
demonstrate here with a proof-of-principle experiment. Moreover, this
means that the advantage of the aforementioned quantum schemes, with a
separation-independent Fisher information, over classical techniques
is smaller than previously thought.

We work with a spatially invariant imaging system and two equally
bright incoherent point sources separated by a distance
$\mathfrak{s}$.  We assume quasimonochromatic paraxial waves with one
specified polarization and one spatial dimension, with $x$ denoting
the image-plane coordinate in direction of the separation.  This
simplified 1D geometry is sufficient to make clear our procedure and
it works for some applications, such as, e.g., spectroscopy. The more
realistic 2D case can be worked along the same lines, although at the
price of dealing with a two-parameter estimation.

If $I(x)$ is the spatial distribution of the intensity in the image from
a point source, commonly called the point-spread function
(PSF)~\cite{Goodman:2004aa}, the direct image can be written as
\begin{equation}
\label{eq:intensity}
p (x|\mathfrak{s}) = \frac{1}{2} 
[I(x-\mathfrak{s}/2)+I(x+\mathfrak{s}/2)] \, ,
\end{equation} 
where $p(x|\mathfrak{s})$ is the probability density for detecting
light at $x$ conditional on the value of $\mathfrak{s}$.  We model
light emission (detection) as a random process (shot
noise)~\cite{Ram:2006aa}.  The precision in estimating $\mathfrak{s}$
is governed by the Fisher information
\begin{equation}
  \label{fisherdef}
  F (\mathfrak{s}) = N \int 
\frac{[ \partial_\mathfrak{s} p(x|\mathfrak{s})]^2} 
  {p(x|\mathfrak{s})} \, dx,
\end{equation}
where $\partial_{\mathfrak{s}}$ is the partial derivative with respect
to $\mathfrak{s}$. In this stochastic scenario, $N$ is the number of
detections, which can be approximately taken as Poissonian with a mean
$N p(x | \mathfrak{s}) dx$. Without loss of generality, we set $N=1$
and evaluate the Fisher information per single
detection. Reintroducing $N$ into the final results is always
straightforward.  The CRLB ensures that the variance of any unbiased
estimator $\hat{\mathfrak s}$ of the quantity $\mathfrak{s}$ is
bounded by the reciprocal of the Fisher information; viz,
$(\Delta \hat{\mathfrak{s}})^{2} \ge 1/F (\mathfrak{s})$.

Since we are chiefly interested in the case of small separations, we
expand $p (x| \mathfrak{s})$ in $\mathfrak{s}$, getting
$ p(x|\mathfrak{s})= I(x) + I^{\prime \prime}(x)/8 \, \mathfrak{s}^{2}
+ O(\mathfrak{s}^4)$, where a prime denotes derivative respect to the
variable.  Observe that the odd powers of $\mathfrak{s}$ make no
contribution, because the contributions from the two PSF components
cancel each other.  The associated Fisher information becomes
\begin{equation}
\label{fishcurse}
F (\mathfrak{s})  =\frac{\mathfrak{s}^2}{16} 
\int \left \{ 
\frac{[I^{\prime \prime}(x)]^2}{I(x)}+O(\mathfrak{s}^2)
\right \} \, dx \, .
\end{equation}
Commuting the order of integration and summation immediately yields a
quadratic behavior for all PSFs: $ F (\mathfrak{s}) \propto
\mathfrak{s}^2+  O(\mathfrak{s}^4)$.
However, such an operation is not always
admissible~\cite{Royden:1988aa}, which leaves room for tempering
Rayleigh's curse with PSF shaping techniques.

\begin{figure}
 \centering 
\includegraphics[width=.95\columnwidth]{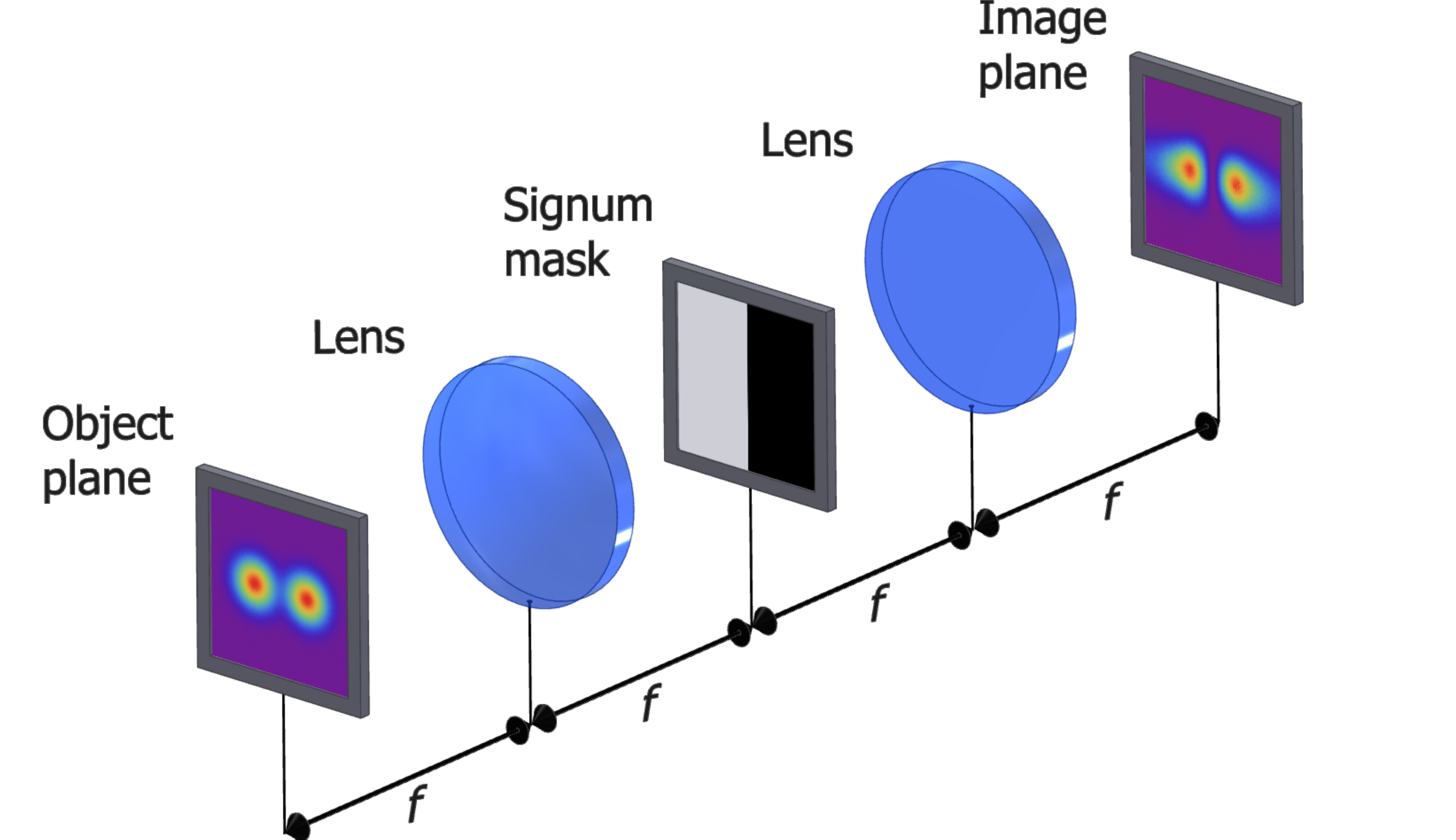}
  \caption{Scheme of an optical coherent $4f$ processor, with a signum
    mask in the Fourier plane.} 
  \label{fig:sketch4f}
\end{figure}

To illustrate this point, let us assume, for the time being, that our
PSF is well approximated by a parabolic profile near the origin; i.e.,
$I(x)\simeq \alpha x^2 $, which implies 
\begin{equation}
\label{eq:parab}
 p(x|\mathrm{s}) \simeq \alpha (x^2+\mathfrak{s}^{2}/4) \, , 
  \qquad x, \mathfrak{s} \ll 1 \, .
\end{equation} 
When this holds true, the integrand in \eqref{fisherdef} reduces to a
Lorentzian function
\begin{equation}
  \label{lorentz}
  F ( \mathfrak{s} ) \simeq  \int 
  \frac{\alpha^2 \mathfrak{s}^2}{4 x^2+\mathfrak{s}^2} \, dx  \, . 
\end{equation}
Because of the strong peak at $x=0$, when $\mathfrak{s} \ll 1$ the
tails of the Lorentzian do not contribute appreciably and can be
ignored. As a result, we get
\begin{equation}
  \label{integrand}
  F (\mathfrak{s}) \simeq  \lambda\, \mathfrak{s} \,,
\end{equation}
with $\lambda = \pi\alpha/2$, and the information is indeed linear
rather than quadratic at small separations.  Note that the proper
normalization of $p(x | \mathfrak{s})$ is guaranteed by higher-order
terms in the expansion \eqref{eq:parab}, but they do not affect the
scaling in \eqref{integrand}.

Next, we show that any PSF can be converted to the form
(\ref{lorentz}) by applying a simple nonabsorbing spatial filter at
the output of the system. In what follows, $\Psi(x)$ indicates the
amplitude PSF, so that $I(x)=|\Psi(x)|^2$, and $\Psi (f) $ its Fourier
transform.  We process the image by a coherent processor, such as,
e.g., a standard $4f$ system~\cite{Goodman:2004aa} schematized in
Fig.~\ref{fig:sketch4f}. In the Fourier plane, each point source gives
rise to $ \Psi (x \pm \mathfrak{s}/2) \mapsto \Psi(f) e^{\pm i \pi f s}$. In
that plane, we apply a signum mask:
$\mathrm{sgn}(f) \Psi(f)e^{\pm i \pi f s}$, where for a real number
$\sgn(t) = |t|/t$ for $t \neq 0$ and $\sgn(0) = 0$.  As the signum is
a pure phase filter, no photons are absorbed.  The signal components
are then convolved with the inverse Fourier transform of the signum
function, which is
$\mathcal{F}^{-1} \{\mathrm{sgn} (f) \} = - i/ (\pi x)$. In this way,
the processor performs
\begin{equation}
  \label{hilbert}
  \Psi^{\mathrm{sgn}}_\pm (x,\mathfrak{s}) = - 
  \frac{i}{\pi}\int  \frac{\Psi(x^{\prime} \pm \mathfrak{s}/2)}
  {x-x^{\prime}} \, dx^{\prime} \, .
\end{equation}
which is the Hilbert transform of the signal.  The optical
implementation of this transform has a long
history~\cite{Kastler:1950aa,Wolter:1950aa,Lohmann:1996aa}. It has
been used in several fields, but most prominently in image processing
for edge enhancement, because it emphasizes the derivatives of the
image.

Applying the change of variable $\xi =x^{\prime}-x$, expanding $\Psi$ to the
second order in the small quantity $x \pm \mathfrak{s}/2$, and using
the spatial symmetry of $\Psi$, we approximate the output amplitudes
after the signum mask by
\begin{equation}
  \Psi^{\mathrm{sgn}}_\pm (x, \mathfrak{s} ) \simeq 
  \frac{i}{\pi}(x \pm \mathfrak{s}/2)
  \int \frac{\Psi^{\prime} (\xi)}{\xi} \, d\xi \,, \qquad x,\mathfrak{s}\ll 1.
\end{equation}

The detection probability density near the origin now takes the
parabolic shape, as discussed before, viz,
\begin{equation}
  \label{eq:exp}
  p^\mathrm{sgn}(x|\mathfrak{s}) =\frac{1}{2} 
  \left [ |\Psi^{\mathrm{sgn}}_-(x,\mathfrak{s})|^2 
    +|\Psi^{\mathrm{sgn}}_+(x,\mathfrak{s})|^2 \right ] 
  \simeq \alpha (x^2+\mathfrak{s}^2/4),
\end{equation}
with $ \alpha= [\int \Psi^{\prime} ( \xi )/\xi \, d\xi ]^2/ \pi^{2}$.
Note carefully that the parabolic behavior (\ref{eq:exp}) is general,
but the value of the coefficient $\alpha$ depends on the explicit form
of the PSF $\Psi (x)$. We thus have a linear Fisher information as in
\eqref{integrand}.  In physical terms, this happens because the
Fourier-space processing incorporates phase information.  In addition,
the combined system consisting of the imaging and PSF reshaping step
remains spatially invariant and so the information about the
separation is not degraded by misaligning the signal and detection
devices, as it happens, for example, when the centroid of the
two-component signal is not perfectly controlled.

\begin{figure}
  \includegraphics[width=.95\columnwidth]{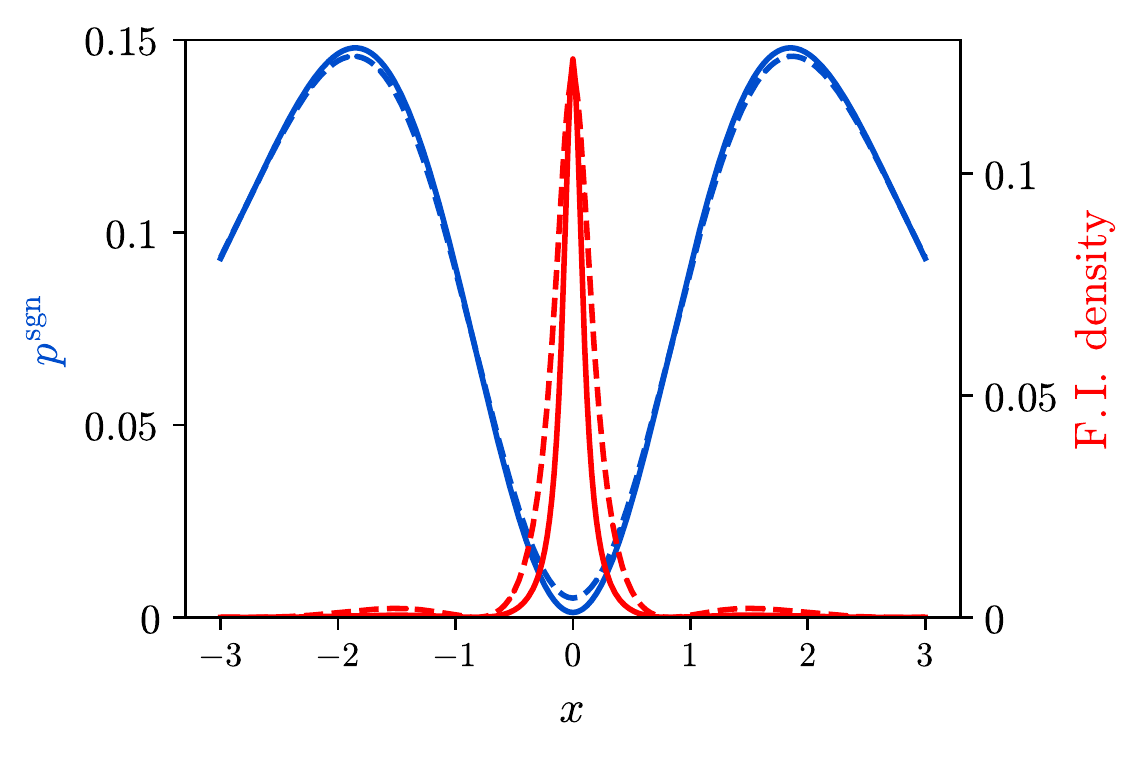}
  \caption{Detection probabilities (blue) and Fisher information
    density [i.e., the integrand in the definition (\ref{fisherdef})]
    (red) corresponding to a Gaussian PSF modified by the
    signum filter for separations $0.2 \sigma$ (solid lines) and
    $0.4 \sigma$ (broken lines). Here, and in all the figures, length is
  in units of $\sigma$.}
  \label{figT1}
\end{figure}

We recall that the Hilbert transform is the essential tool for getting
the dispersion relations~\cite{King:2009aa}, which relate the real and
imaginary parts of the response function (i.e., susceptibility) of a
linear causal system.  If we think of $\Psi(x)$ as an absorption
profile near resonance , we realize that the real part of
susceptibility shows anomalous dispersion---linear slope---near the
resonance: after squaring we then get a parabolic $p(x, \mathfrak{s})$
near the origin.

Let us ellaborate our proposal with the relevant example of a system
characterized by a Gaussian PSF:
$ \Psi(x)= (2\pi\sigma^2)^{-1/4} \exp (- \tfrac{1}{4} x^2/\sigma^2 )$,
where $\sigma$ is an effective width that depends on the wavelength.
Henceforth, we take $\sigma$ as our basic unit length, so the
corresponding magnitudes (such as separation, variance, etc) appear as
dimensionless. Apart from its computational efficiency, the Gaussian
PSF approximates fairly well the Airy distribution when the
illumination is done by a Gaussian distribution that apodises the
circular aperture.

\begin{figure}
\centerline{\includegraphics[width=0.95\columnwidth]{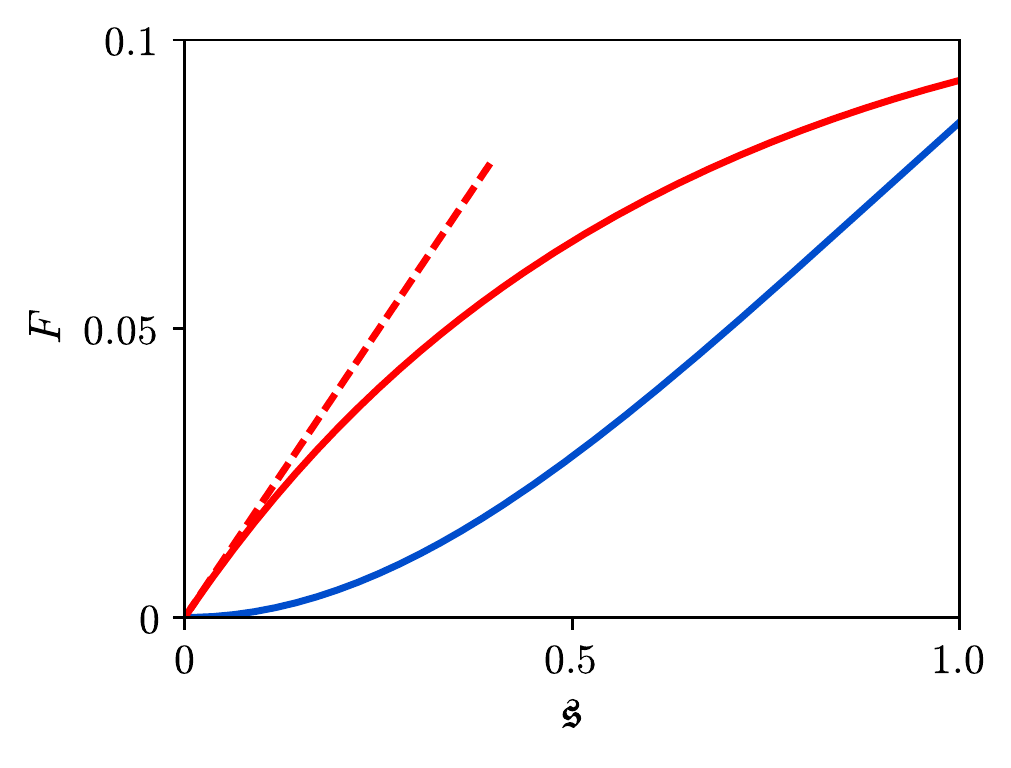}}
\caption{Fisher information about separation for imaging with a Gaussian
  PSF with (red concave line) and without (blue convex line) the
  signum filter. The asymptotic behavior of the superresolution given
  by the right hand side of~\eqref{asymptot} is also shown (broken
  red line). 
\label{figT2}}
\end{figure}

The Fisher information associated with the direct imaging is obtained
from \eqref{fisherdef}; the result reads
\begin{equation}
  F_\mathrm{direct} (\mathfrak{s}) \simeq
  \frac{(\mathfrak{s}/\sigma)^2}{8\sigma^2} \, ,
\end{equation}
confirming once again the quadratic scaling of Rayleigh's curse.  This
is to be compared with the information accessible by signum-filter
enhanced detection.  We first perform the Hilbert transform of the
Gaussian PSF;
\begin{equation}
  \label{dawson}
  |\Psi^\mathrm{sgn}_\pm(x,\mathfrak{s})|^2 =
  \frac{2\sqrt{2}\; D \left( \frac{x\pm\mathfrak{s}/2}{2\sigma}\right)^2}
  {\pi^{3/2}\sigma} \, ,
\end{equation}
where $D (z)$ denotes Dawson's integral~\cite{Temme:2010aa}. Similar
results have been reported for the dispersion relations of a Gaussian
profile~\cite{Wells:1999aa}.  In particular, $D(- z) = - D(z)$ and
$D(z) \simeq z (1- \tfrac{2}{3}z^{2})$ for $z \rightarrow 0$, so the
dominant behavior is indeed linear. Therefore, the expansion in
\eqref{eq:exp} holds with $\alpha=(2\pi^3)^{-1/2}\sigma^{-3}$ and, in
consequence,
\begin{equation}
  \label{asymptot}
  F^\mathrm{sgn} (\mathfrak{s}) \simeq \frac{(\mathfrak{s}/\sigma)}
  {2\sqrt{2 \pi} \sigma^2} \, .
\end{equation}

The detection probabilities and Fisher information densities typical
for a signum-enhanced detection with a Gaussian PSF are shown in
Fig.~\ref{figT1} for two different values of $\sigma$.  Notice that
nonzero separation is evidenced by nonzero readings at the center of
the image. Interestingly, most of the information on the separation
comes from detections near the origin and this region shrinks with
decreasing $\mathfrak{s}$.

\begin{figure}[tbp]
  \centering
  \includegraphics[width=0.95\columnwidth]{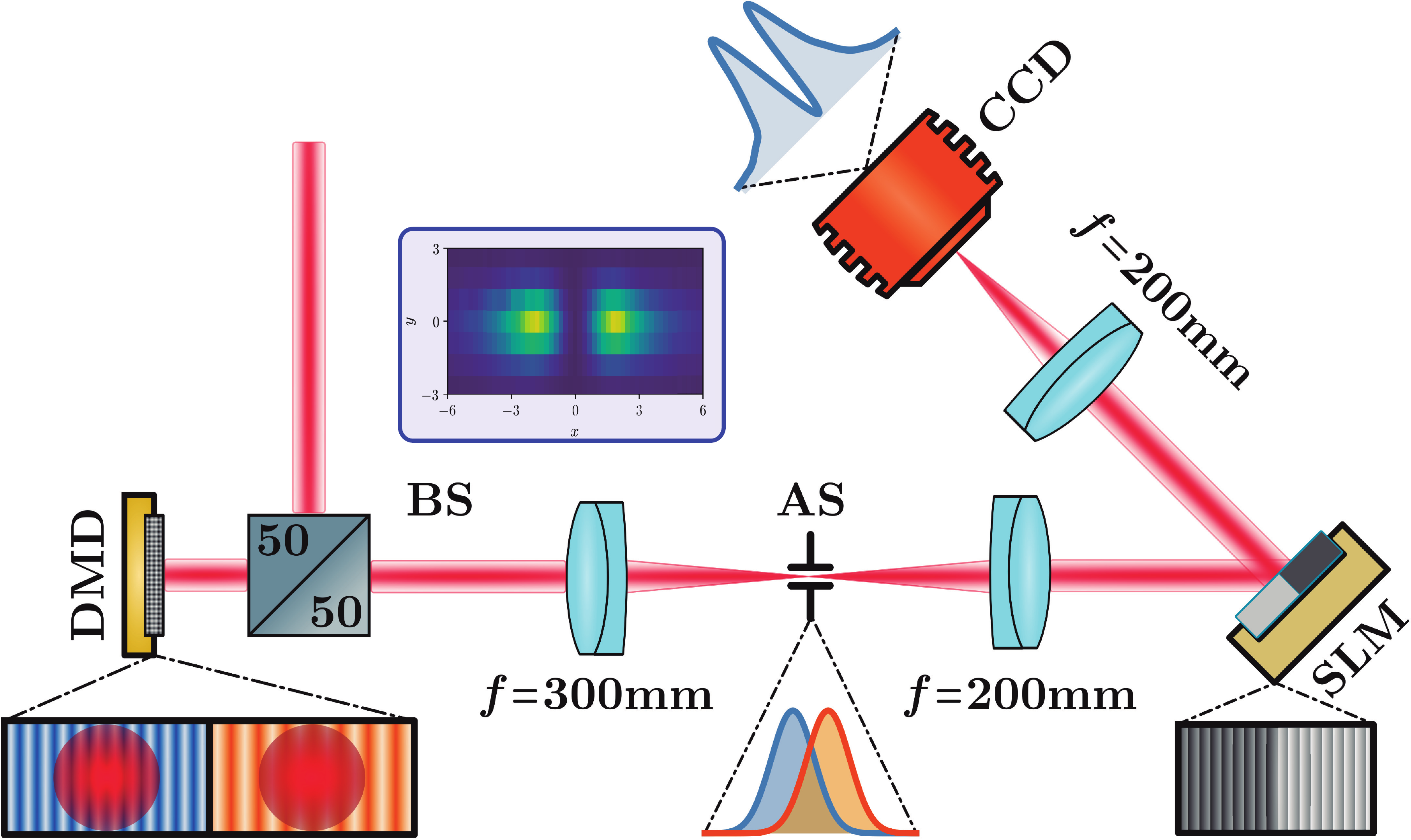}
  \caption{Experimental setup. The notations used are as follows: DMD:
    digital micromirror chip, BS: beam-splitter, AS: aperture stop at
    the Fourier plane of the lens, and SLM: spatial light
    modulator. In the inset, we show a typical intensity scan recorded
    with zero separation setting and a total detection count of
    $434,000$. The separation is estimated from the total number of
    detections registered in the central pixel column.}
  \label{fig:setup}
\end{figure}

The superresolution potential of our technique is illustrated in
Fig.~\ref{figT2}.  The linear scaling can provide big advantages in
terms of the resources required to measure very small separations. For
example, to measure a $10\times$ smaller separation to a given
precision requires $100\times$ more detection events with the
conventional setup, while just $10\times$ would do with the new
technique. For a fixed photon flux this translates into shorter
detection times. At the same time, the new technique is simple to
implement in existing imaging devices, such as telescopes, microscopes
or spectrometers.

We have implemented the method with the setup sketched in
Fig.~\ref{fig:setup}.  Two mutually incoherent equally bright point
sources were generated with a controlled separation.  After
preparation, this signal was detected using a signum spatial-frequency
filter, by which the original Gauss PSF is reshaped into the Dawson
form as in~\eqref{dawson}.

A spatially coherent, intensity-stabilised Gaussian beam was used to
illuminate a digital micro-mirror chip (DMD, Texas Instruments) with a
mirror pitch of $10~\mu$m. Two sinusoidal grating patterns with very
close spatial frequencies were created by the DMD, which allows for a
very precise control of the angular separation in a chosen diffraction
order. Angular separations as small as $4.6~\mu$rad were realized;
these correspond to a linear separation of 0.042$\sigma$.  To ensure
incoherence, one pattern was ON at a time, while keeping the switching
time well below the detector time resolution.  Imaging with an
objective of focal length $f=300$~mm gave rise to two spatially
separated Gaussian spots of $\sigma=33.2~\mu$m. An aperture stop was
used to cut-off unwanted diffraction orders.  This completes the
direct imaging stage.

In the signum-enhanced imaging part, a phase spatial light modulator
(SLM) (Hammamatsu) with square pixels of $20 \times 20~\mu$m was
operated in the Fourier plane of a standard $4f$ optical system. The
SLM implemented the signum mask hologram calculated as an interference
pattern between a phase unit-step and a blaze grating, allowing for
over $0.9$ transfer efficiency. Finally, the output signal was
measured by a CCD camera (Basler) with $7.4\times 7.4~\mu$m pixels
positioned at the output of the $4f$ processor. Vertical $4$-pixel
binning was activated to reduce the readout noise, so the effective
pixel size is $7.4 \times 29.6~\mu$m. The corresponding signals used
in the reconstruction process, resulting from summing three pixels,
were in the range of 120--253 photoelectrons, in comparison to a sum
of $3 \times 7$ photoelectron readout noise. The camera exposure time
was set to $100$~ms to keep the dark noise contribution negligible.

\begin{figure}
\centerline{\includegraphics[width=.95\columnwidth]{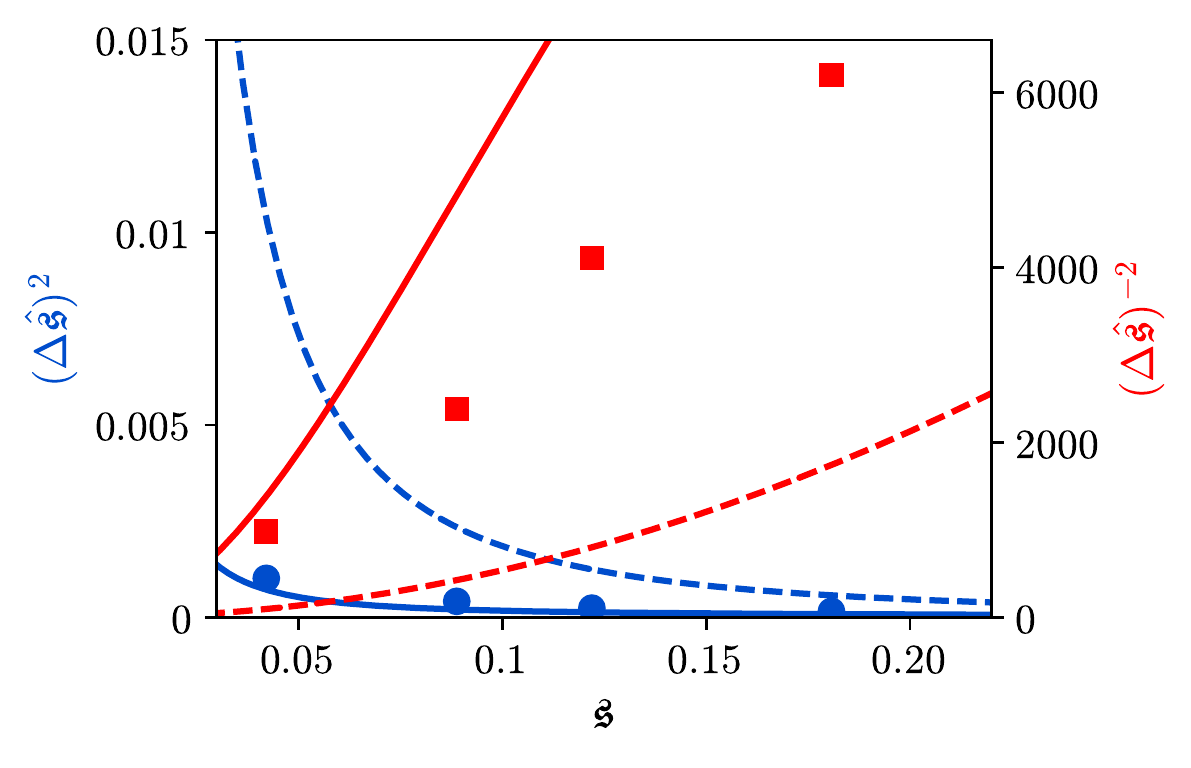}}
\caption{Experimental variances of the separation estimator
  (blue dots) compared with the direct detection  (blue
  broken line) and the signum-enhanced limit (solid blue
  line). The latter is corrected for the finite pixel size of
  $7.6~\mu$m. For completeness, the reciprocal of the variances,
  called the precisions, are shown in red.}
 \label{figE1}
\end{figure}

Several separations, ranging from about $0.042\sigma$ ($1.4~\mu$m) to
$0.18\sigma$ ($6~\mu$m), were measured. Two hundred intensity scans
were recorded for each separation setting. One typical 2D scan is
shown in the inset of Fig.~\ref{fig:setup}. Since the two incoherent
points are separated horizontally, no information about separation is
lost by collecting pixel counts column-wise. The resulting 1D
projections are samples from the theoretical intensity distribution
$p^{\mathrm{sgn}} (x,\mathfrak{s})$, see Fig.~\ref{figT1}.  Notice
that for small separations only the central parts of the projections
contribute significant information. In particular all pixel columns,
except the central one, can be ignored in the raw data in the inset of
Fig.~\ref{fig:setup}.  Therefore, each 2D intensity scan is reduced to
a single datum---the total number of detections in the central pixel
column.

We express the response of the real measurement by a second-order
polynomial on the separation
$p^{\mathrm{}} (\mathfrak{s}) = a + b \mathfrak{s}^2$ and estimate the
coefficients from a best fit of the mean experimental detections.  For
each separation, we calculate the estimator mean
$\langle\widehat{\mathfrak{s}}\rangle$ and
variance$(\Delta \widehat{\mathfrak{s}})^2$.

\begin{figure}
  \centerline{\includegraphics[width=.90\columnwidth]{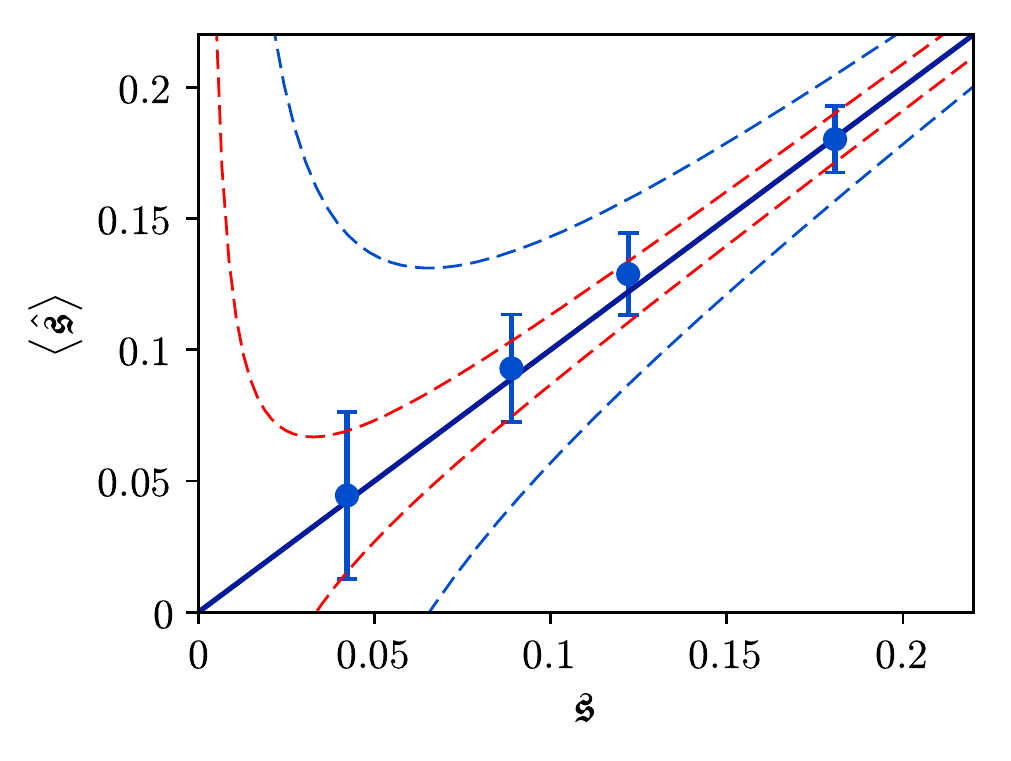}}
  \caption{Estimation of the separation from signum-enhanced
    imaging. Estimator means (dots) and standard deviations (error
    bars) are shown. Same statistics is provided for the best unbiased
    estimators from direct (blue lines) and signum-enhanced (red
    lines) imaging as given by the CRLB. The latter takes in account
    the finite pixel size used in the experiment.}
  \label{figE2}
\end{figure}

Experimental results are summarized in Figs.~\ref{figE1} and
\ref{figE2}. Figure~\ref{figE1} compares the experimentally determined
variances with the theoretical limits of the direct and
signum-enhanced imaging for a Gaussian PSF and $434,000$ detections
per measurement.  Reciprocal quantities (precisions) are also
shown. Signum-enhanced imaging clearly breaks the quadratic
Rayleigh's curse in the whole range of measured separations, with an
variance improvements up to $10\times$ compared with the direct
imaging.  Notice also the apparent linear behavior of experimental
precisions (red symbols) as compared to the quadratic lower bound
predicted for the direct imaging (red broken line).

More estimator statistics is shown in Fig.~\ref{figE2}. Experimental
estimates are nearly unbiased and not much worse than the theoretical
limit calculated for the finite pixel size used in the
experiment. Engineering the PSF brings about reliable separation
estimates in the region where direct imaging fails, as for example for
separations $\mathfrak{s} \lessapprox 0.07\sigma$ in Fig.~\ref{figE2}.

In summary, we have demonstrated a robust experimental violation of
Rayleigh's curse. Experimental imperfections prevent one from achieving
the ultimate limit shown in Fig.~\ref{figT2}.  For larger separations, 
systematic errors and setup instability make important contributions
to the total (small) error. For very small separations, the measured
signal is very weak and background noise becomes the limiting
factor. Further improvements are possible by optimizing the noise
statistics and resolution of the camera.

Finally, one could wonder whether a different filter could yield a
better scaling of the Fisher information using direct imaging. The
dispersion relations suggest that this behavior is largely determined
by the zeros of the PSF. Additional work is needed to explore all
these issues, but the simplicity of the signum mask makes it very
attractive for superresolution applications.

We acknowledge financial support from the Grant Agency of the Czech
Republic (Grant No.~18-04291S), the Palack\'y University (Grant No.
IGA\_PrF\_2018\_003), the European Space Agency's ARIADNA scheme, and
the Spanish MINECO (Grant FIS2015-67963-P).


\begin{thebibliography}{10}
\newcommand{\enquote}[1]{``#1''}

\bibitem{Kay:1993aa}
S.~M. Kay, \emph{Fundamentals of Statistical Signal Processing}, vol.~1
  (Prentice Hall, 1993).

\bibitem{Dekker:1997aa}
A.~J. den Dekker and A.~van~den Bos, J. Opt. Soc. Am. A \textbf{14}, 547 (1997).

\bibitem{Hemmer:2012aa}
P.~R. Hemmer and T.~Zapata, J. Opt. \textbf{14}, 083002 (2012).

\bibitem{Tsang:2016aa}
M.~Tsang, R.~Nair, and X.-M. Lu, Phys. Rev.~X \textbf{6}, 031033 (2016).

\bibitem{Nair:2016aa}
R.~Nair and M.~Tsang, Phys. Rev. Lett. \textbf{117}, 190801 (2016).

\bibitem{Nair:2016ab}
R.~Nair and M.~Tsang, Opt. Express \textbf{24}, 3684 (2016).

\bibitem{Tsang:2017aa}
M.~Tsang, New J. Phys. \textbf{19}, 023054 (2017).

\bibitem{Petz:2011aa}
D.~Petz and C.~Ghinea, \emph{Introduction to {Q}uantum {F}isher {I}nformation}
  (World Scientific, 2011), pp. 261--281.

\bibitem{Lupo:2016aa}
C.~Lupo and S.~Pirandola, Phys. Rev. Lett. \textbf{117}, 190802 (2016).

\bibitem{Rehacek:2017aa}
J.~Rehacek, M.~Pa{\'u}r, B.~Stoklasa, Z.~Hradil, and
L.~L. S{\'a}nchez-Soto, Opt. Lett. \textbf{42}, 231 (2017).

\bibitem{Paur:2016aa}
M.~Paur, B.~Stoklasa, Z.~Hradil, L.~L. Sanchez-Soto, and J.~Rehacek, 
Optica  \textbf{3}, 1144 (2016).

\bibitem{Yang:2016aa}
F.~Yang, A.~Taschilina, E.~S. Moiseev, C.~Simon, and A.~I. Lvovsky, Optica
  \textbf{3}, 1148 (2016).

\bibitem{Yang:2017aa}
F.~Yang, R.~Nair, M.~Tsang, C.~Simon, and A.~I. Lvovsky, Phys. Rev. A
  \textbf{96}, 063829 (2017).

\bibitem{Tham:2016aa}
W.~K. Tham, H.~Ferretti, and A.~M. Steinberg, Phys. Rev. Lett. \textbf{118},
  070801 (2016).

\bibitem{Goodman:2004aa}
J.~W. Goodman, \emph{Introduction to {F}ourier {O}ptics} (Roberts and
Company, 2004).

\bibitem{Ram:2006aa}
S.~Ram, E.~Sally~Ward, and R.~J. Ober,
PNAS \textbf{103}, 4457 (2006).

\bibitem{Royden:1988aa}
H.~L. Royden, \emph{Real Analysis} (Prentice Hall, 1988).

\bibitem{Kastler:1950aa}
A.~Kastler, Rev. Opt. \textbf{29}, 308 (1950).

\bibitem{Wolter:1950aa}
H.~Wolter, Ann. Phys. \textbf{7}, 341 (1950).

\bibitem{Lohmann:1996aa}
A.~W. Lohmann, D.~Mendlovic, and Z.~Zalevsky, Opt. Lett. \textbf{21}, 281 (1996).

\bibitem{King:2009aa}
F.~W. King, \emph{Hilbert transforms} 
(Cambridge University Press,  2009).

\bibitem{Temme:2010aa}
N.~M. Temme, \emph{NIST Handbook of Mathematical Functions} (Cambridge
  University Press, 2010), chap. Error Functions, Dawson's and Fresnel
  Integrals.

\bibitem{Wells:1999aa}
R.~J. Wells, J. Quant. Spect. Rad. Transfer \textbf{62}, 29 (1999).

\end{thebibliography}

\end{document}